\documentclass[showpacs,notitlepage,floatfix,citeautoscript,superscriptaddress,prb,reprint]{revtex4-2}
\usepackage[english]{babel}
\usepackage{graphicx}% Include figure files
\usepackage{dcolumn}% Align table columns on decimal point
\usepackage{bm}% bold math
\usepackage{hyperref}% add hypertext capabilities
\usepackage[dvipsnames,svgnames,table]{xcolor}

\usepackage{color,soul}
\usepackage{wasysym}
\usepackage{mathrsfs}
\usepackage{times}
\usepackage[T1,OT1]{fontenc}
\usepackage{fancyhdr}
\usepackage{float}

\hypersetup{
pdfstartview={FitH},% fits the width of the page to the window
colorlinks=true,    % false: boxed links; true: colored links
linkcolor=NavyBlue, % color of internal links
citecolor=Maroon,   % color of links to bibliography
filecolor=NavyBlue, % color of file links
urlcolor=NavyBlue   % color of external links
}

\begin{document}

\newcommand{\JM}[1]{\textcolor{red}{#1}}

\title{Inducing a topological transition in graphene nanoribbons superlattices by external strain}

\author{E. Flores}
\affiliation{Departamento de Física, Facultad de Ciencias Físicas y Matemáticas, Universidad de Chile, Santiago, Chile}

\author{Jos\'e D. Mella}
\affiliation{Departamento de Física, Facultad de Ciencias Físicas y Matemáticas, Universidad de Chile, Santiago, Chile}

\author{E. Aparicio}
 \affiliation{CONICET and Universidad de Mendoza, Mendoza, 5500, Argentina}
 
\author{R. I. Gonzalez}
 \affiliation{Centro de Nanotecnología Aplicada, Facultad de Ciencias, Universidad Mayor, Santiago 8580745, Chile.} 
 \affiliation{Centro para el Desarrollo de la Nanociencia y la Nanotecnolog\'ia, CEDENNA, Santiago, Chile}

\author{C. Parra}
\affiliation{Laboratorio de Nanobiomateriales, Departamento de Física, Universidad Técnica Federico Santa María, Valparaiso, Chile}

\author{E.M. Bringa}
 \email{ebringa@yahoo.com}
 \affiliation{CONICET and Universidad de Mendoza, Mendoza, 5500, Argentina}
 \affiliation{Centro de Nanotecnología Aplicada, Facultad de Ciencias, Universidad Mayor, Santiago 8580745, Chile.}

\author{F. Munoz}
\email{fvmunoz@gmail.com}
 \affiliation{Departamento de F\'isica, Facultad de Ciencias, Universidad de Chile, Santiago, Chile}
 \affiliation{Centro para el Desarrollo de la Nanociencia y la Nanotecnolog\'ia, CEDENNA, Santiago, Chile}

\date{\today}
\begin{abstract}
Armchair graphene nanoribbons, when forming a superlattice, can be classified in different topological phases, with or without edge states. By means of tight-binding and classical molecular dynamics  (MD) simulations, we studied the electronic and mechanical properties of some of these superlattices. MD shows that fracture in modulated superlattices is brittle, as for unmodulated ribbons, and that occurs at the thinner regions, with staggered superlattices achieving a larger fracture strain.
We found a general mechanism to induce a topological transition with strain, related to the electronic properties of each segment of the superlattice, and by studying the sublattice polarization we were able to characterize the transition and the response of these states to the strain. For the cases studied in detail here, the topological transition occurred at $\sim$3-5 \% strain, well below the fracture strain.
The topological states of the superlattice -if present- are robust to strain even close to fracture. Unlike the zero-energy edge states found in the zig-zag edges of graphene nanoribbons, the superlattice states shows signatures of being particularly insensitive to disorder, even in real space. 

\end{abstract}

\maketitle

\section{\label{sec:intro}Introduction}

Since its exfoliation, graphene has been an object of intense research by its exotic electronic properties, which are interesting for basic and applied science\cite{kane2005quantum,zhang2005experimental,novoselov2004electric,novoselov2005two,san2009pseudospin,xia2010,wang2008gate,xia2009ultrafast}. Nevertheless, the lack of a bandgap has hindered its utility in electronic devices and the realization of exotic topological states. Different approaches, experimental and theoretical, to generate gapped graphene are developed, such as graphene bilayer with a gate voltage \cite{xia2010,zhang2009} or applying uniaxial \cite{ni2008uniaxial,pereira2009tight} as well as topological currents\cite{sui2015,ju2015,munoz2016bilayer}. Even more exotic topological effects are predicted to appear in the twisted bilayer graphene\cite{park2021higher}.

Another option to develop a bandgap in graphene-based materials is the usage of graphene nanoribbons (GNRs), which according to their chirality may be insulating or metallic\cite{son2006energy}. However, the traditional synthesis of GNRs offers little control over chirality and uniformity. The recent development of a bottom-up approach to synthesize GNR from precursor molecules has allowed precise control of chirality and width of the GNRs. Among them, the most promising for applications are the armchair GNR, denoted as $N$-AGNRs, with $N$ the number of C atoms along the width of the ribbon\cite{jacobse2017electronic,el2020controlled,Houtsma21,Sun20adma,Lawrence20,zdetsis2021topological}. This bottom-up approach has been successfully applied to build more complex systems,\cite{Ortiz18} such as GNRs with a specific periodic modulation in their width.\cite{Sun2021} Recently, two works by Gr\"onig \textit{et al.}\cite{groning2018} and Rizzo \textit{et al.}\cite{rizzo2018} demonstrated that periodic superlattices of $N$-AGNR/$M$-AGNR (different width) host topologically protected edge states, in close analogy to the Su-Schieffer-Heeger (SSH) model of polyacetylene\cite{SSH}. % Indeed, $N$-AGNRs can be topologically classified by their width. 
Indeed, there is a width dependence in the topological classification of $N$-AGNR, where topological states appear at the boundaries between two insulating AGNRs with different topological phase\cite{Cao2017Topo,Lee2018,Kuan17,Jiang2021}. An engaging system that meet with this condition is the 7-AGNR/9-AGNR interface\cite{groning2018,rizzo2018}.

The finite size plays an important role in $N/M$-AGNR (with different topological phase) interfaces, where two of these boundary states are close enough to interact and hybridize with each other. Additionally, if these boundary states are periodically repeated, they form a superlattice energy band. By controlling the pattern of this periodicity, it is possible to have two different interaction strengths between different topology-derived localized states. In this way, the SSH model can be emulated in the low-energy description of modulated AGNRs. The SSH model has two different topological phases, and they translate -by the \textit{bulk-boundary correspondence} principle- in the existence/absence of zero-energy, localized edges states at the end of the sample\cite{Rhim18,pendas2019chemical}. In the case of a superlattice, these edge states will appear at the superlattice's edges\cite{munoz2018topological}.

In this contribution, we show \textit{(i)} the resilience of the superlattice topological edge states even under a very large strain, since these states show an special robustness derived from the superlattice geometry, and \textit{(ii)} the possibility of inducing a topological phase transition in the modulated AGNR superlattice. Our focus is in the elastic region under strain, but we will include some discussion about the system's fracture process. We will start, in Sec.~\ref{sec:theory} by setting the notation and making a quick summary of the theory behind the band-engineering in the width-modulated AGNRs. In Sec.~\ref{sec:mech} we will show the mechanical properties of the modulated AGNR under strain. Afterwards, in Sec.~\ref{sec:elec}, we will discuss the electronic properties of the AGNR superlattices under strain. The topological transition and the robustness of the states will be discussed in Sec.~\ref{sec:slp} in light of the sublattice polarization of ghraphene-like systems. In Sec.~\ref{sec:methods} we will elaborate on the methods used in our study.

\section{AGNR superlattices}
\label{sec:theory}

According to the their width $N$, an AGNR can be classified in one of three different families, $N=\{3p, 3p+1, 3p+2\}$ (where $p$ is a positive integer)\cite{son2006energy}. According to their Zak phase, the families $N=3p$ and $N=3p+1$ belong to a different topological class\cite{Cao2017Topo}, while the $N=3p+2$ is predicted to be metallic (excluding a small gap opening due to finite-width effect). Indeed, the one-dimensional $k$-points of a  $N=3p+2$  AGNR passes over the $K$ and $K'$ points of graphene (graphene's Dirac cones location), explaining its lack of a band-gap.\cite{datta2005quantum} In the other families of AGNRs the band-gap decreases with $N$.

At the interface between $N=3p$, $N'=3p+1$ AGNRs, due to their different topological invariant,\cite{Cao2017Topo} a localized state within the band-gap must appear (usually called a `topological' state). For instance, Rizzo \textit{et al.} found these states in the interface between 7-AGNR and 9-AGNR.\cite{rizzo2018} The penetration of these states into the AGNR bulk is inversely proportional to the system band-gap, \textit{e.g.} the topological state penetrates deeper into a 9-AGNR region than into the 7-AGNR. In the case of a superlattice of AGNRs of different widths, these localized states hybridize with each other, forming extended states that conserve some of the properties of their constituent localized states. Generally, the localized states penetrate deeper into one region than into the other, allowing the formation of a bipartite lattice in the low-energy description (\textit{i.e.} analogous to a diatomic chain), even if each AGNR segment has the same length. 
The physics of such low-energy description is captured by the SSH model, which has two gapped phases, trivial and topological. Notably, one of them predicts the existence of topologically protected edge states. Therefore, in the superlattice of different AGNRs, the edge states are made from SSH-like topologically derived interface states or superlattice states.\cite{munoz2018topological} In this work we extend the work by Gr\"oning \textit{et al.} including strain, which can modify topological states.

\begin{figure}[h]
    \centering
    \includegraphics[width=\columnwidth]{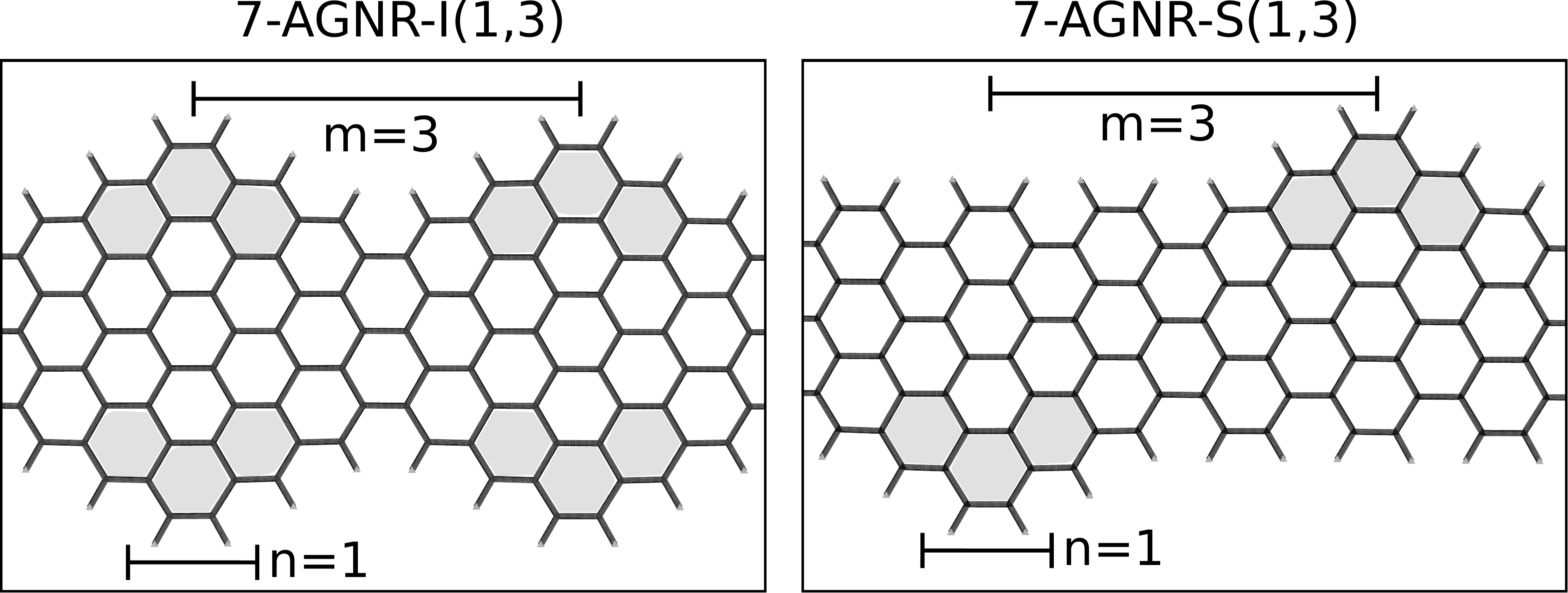}
    \caption{Schematic of the $N$-AGNR-I($n,m$) and $N$-AGNR-S($n,m$) superlattices. $N$ is the width in C atoms of the  \textit{backbone}, \textit{i.e.} removing the shaded regions. The shaded regions have a width $N'=N+4$ for the $I$ (inline), and $N'=N+2$ for the $S$ (staggered) conformations. The indexes $n,m$ are the length of the wider region and the distance (in unit cells of the backbone AGNR) between  adjacent wider regions, respectively. The panel of the $I$($S$) system shows two (one) unit cell.}
    \label{fig:scheme}
\end{figure}

In the following, we will use Gr\"oning \textit{et al.} notation.\cite{groning2018} The modulated AGNRs are labeled as $N$-AGNR-$S$($n,m$), or $N$-AGNR-$I$($n,m$) where $N$ stands for the backbone width, $n$ is the length of the wider region, $m$ is the distance between the wider regions, and $S$ or $I$ defines the ribbon as `staggered' or `inline'. The wider regions has width $N'=N+2$ and $N'=N+4$ for the  $S$ and $I$ cases, respectively. See Fig~\ref{fig:scheme}.

\begin{figure}[h]
    \centering
    \includegraphics[width=\columnwidth]{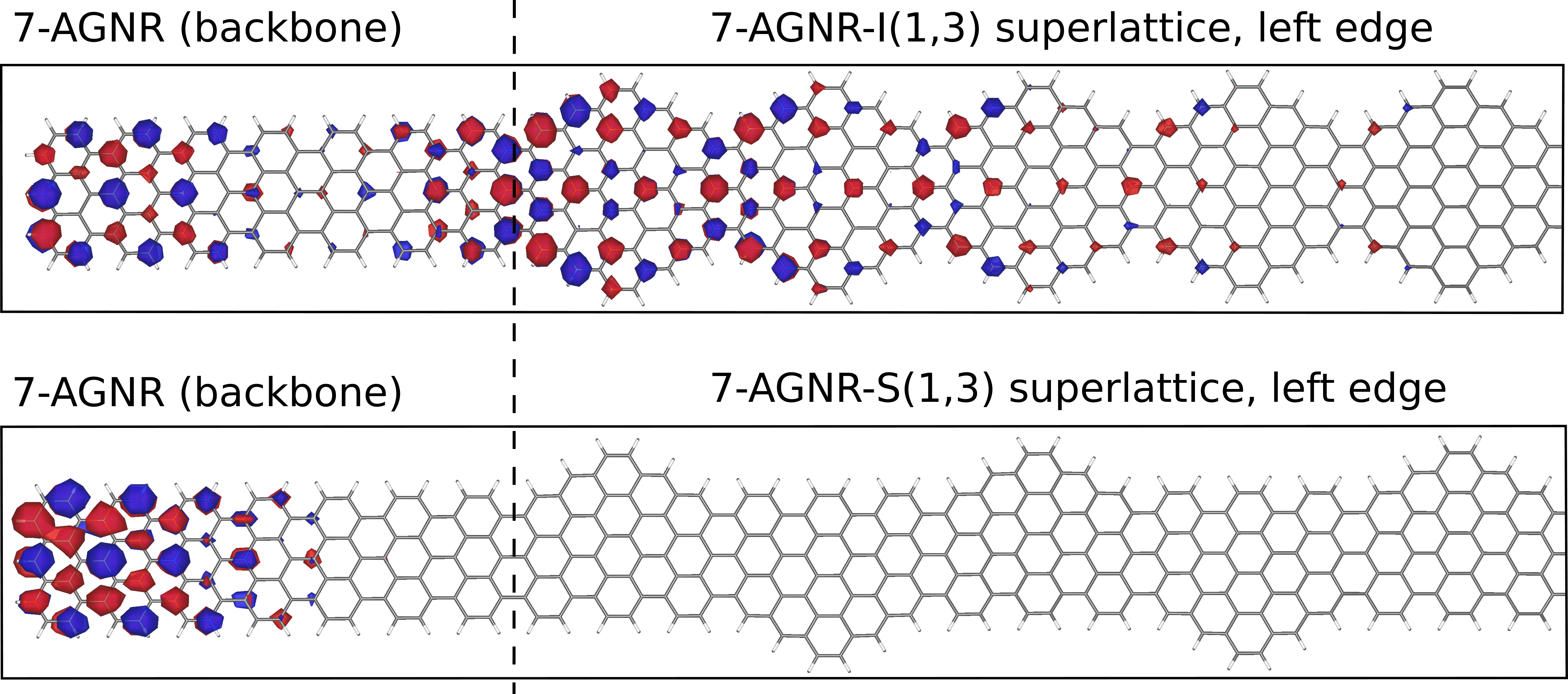}
    \caption{Model of the left edge of finite superlattices 7-AGNR-$I$(1,3) and 7-AGNR-$I$(1,3). The dashed line marks the border of the superlattices, connected to finite 7-AGNR regions. The whole system is much larger, and only the left edge is shown. The wave-functions of the topological states are shown. The  7-AGNR-$I$(1,3) and 7-AGNR-$S$(1,3) have a topological and a trivial phase, respectively, reflected in the existence or lack of localized states at the superlattice edge (dashed line). At the edge of the 7-AGNR there is  another topological state, not related to the superlattice. In 7-AGNR-$I$(1,3), both states form a linear combination, and only one of them is shown.}
    \label{fig:wf-scheme}
\end{figure}

In the remainder of this article, we will restrict to just two different AGNR superlattices, 7-AGRN-$I$(1,3) and 7-AGNR-$S$(1,3). Recently, both systems were experimentally studied by Gr\"oning \textit{et al.}\cite{groning2018} where the low-energy states were described by a superlattice realization of the SSH model. While 7-AGNR-$S$(1,3) is in the trivial phase of the SSH model, 7-AGNR-$I$(1,3) has a non-trivial phase resulting in localized states at its edges. These edge states are protected by the chiral symmetry of the Hamiltonian (\textit{i.e.} they should be degenerated as long as the on-site energy of each interface of the superlattice is the same), and they are sublattice polarized (see Fig.~\ref{fig:wf-scheme}). 

Additionally to the possible presence of the SSH-like topological states (derived from the superlattice geometry), the AGNRs have other topological states: they come from the sublattice split zig-zag borders, and they are not related to the superlattice geometry. To distinguish between both types of topological states, we follow the approach of Gr\"oning \textit{et al.} using an extended backbone, without the modulation. Fig.~\ref{fig:wf-scheme} shows this type of edge, while in both cases, the zig-zag border has topological states, only in 7-AGNR-$I$(1,3) there are topological states at the end of the superlattice. At the Fermi energy of this system, four states exist with almost the same energy. Two are at the (left and right) superlattice boundary, and another two at the backbone's zig-zag edge. These states have a residual interaction, and they hybridize due to finite-size effects.

\section{Mechanical Properties}
\label{sec:mech}

Mechanical properties of any material become relevant every time a technological application is intended, since mechanical deformation can lead to changes in electronic properties, and plastic deformation and fracture can limit the range of applicability of the material.\cite{gao2016unusual} In this section we analyze the mechanical response of 7-AGNR-$S$(1,3) and 7-AGNR-$I$(1,3) structures to a longitudinal uniaxial tension, and compare the results with the same test on 7-AGNR and 11-AGNR samples, which represent the narrowest and the widest versions of uniform width AGNR. We use classical molecular dynamics (MD), and details are given in section \ref{sec:methods}, and in the Supplementary Material, Figs. \ref{fig:AGNRS} and \ref{fig:AGNRI}.

\begin{figure}
    \centering
    \includegraphics[width=1\columnwidth]{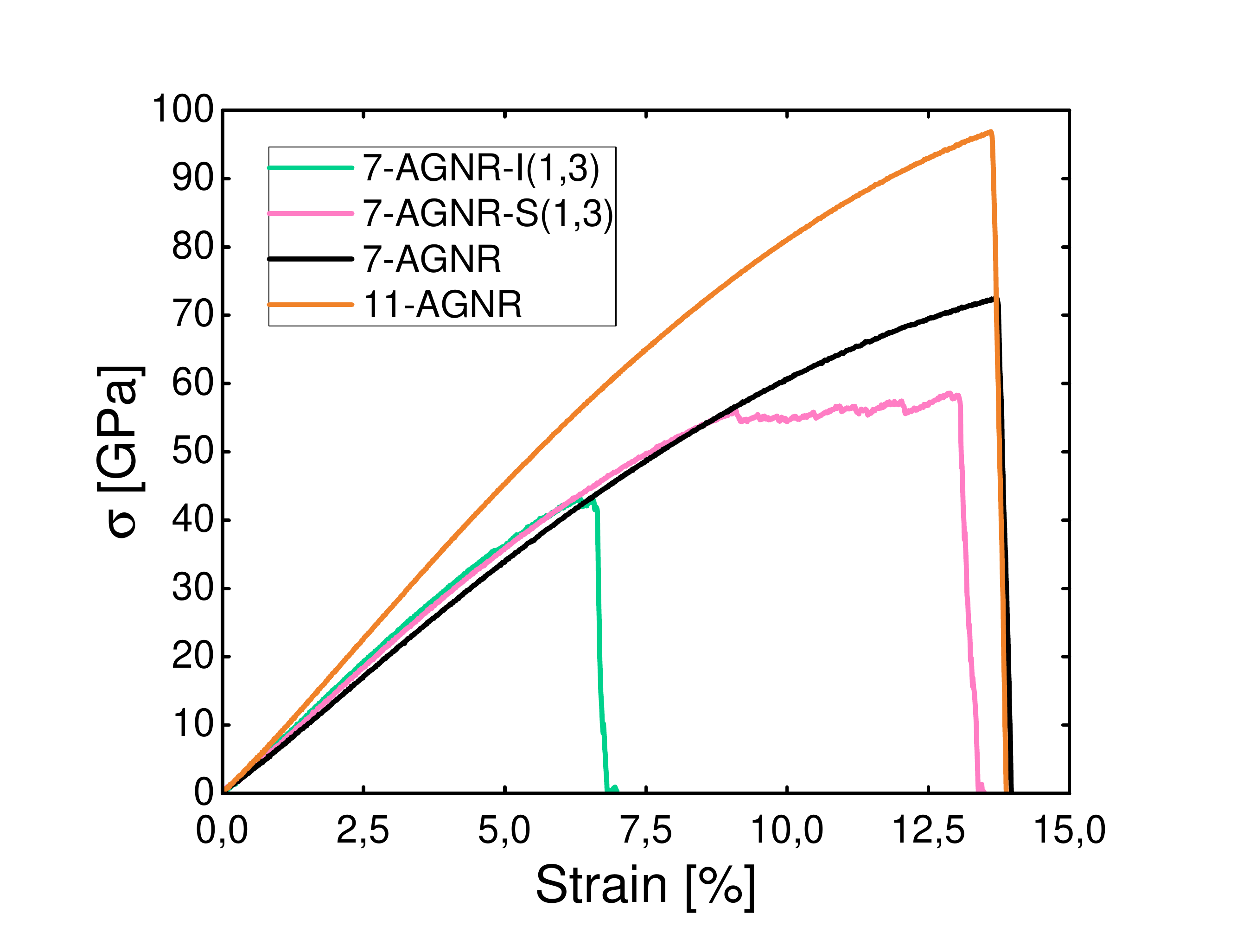}
    \caption{Stress-Strain curves for uniaxial deformation of the 7-AGNR-$S$(1,3) and 7-AGNR-$I$(1,3) superlattices. Results for the 7-AGNR and 11-AGNR are also shown.}
    \label{fig:strain-stress}
\end{figure}

Figure~\ref{fig:strain-stress} displays the Stress-Strain curves for all four cases. The unmodulated AGNR show the typical behavior seen for GNR \cite{aparicio2020simulated}: an initial linear elastic region, followed by a non-linear elastic region, ending in brittle fracture. The elastic module depends on ribbon width \cite{aparicio2020simulated}, but they both fracture near 13.8\% strain. The elastic modulus of both modulated 7-AGNR is similar to the one for the 7-AGNR unmodulated ribbon. The 7-AGNR-$I$(1,3) case is similar overall to a typical AGNR, but modulation leads to a much lower fracture strain, at 6.7\%. The 7-AGNR-$S$(1,3) displays an extended plastic region. This is because bonds between nanoribbon sections that give modulation are not broken, but extremely weak because they are stretched beyond 0.2 nm, which is the usual bond-breaking length for C nanostructures \cite{tangarife2019molecular}. This allows the fracture strain to reach a value of 13.1\%, close to the one for the unmodulated ribbons.

\begin{figure*}
    \centering
    \includegraphics[width=2\columnwidth]{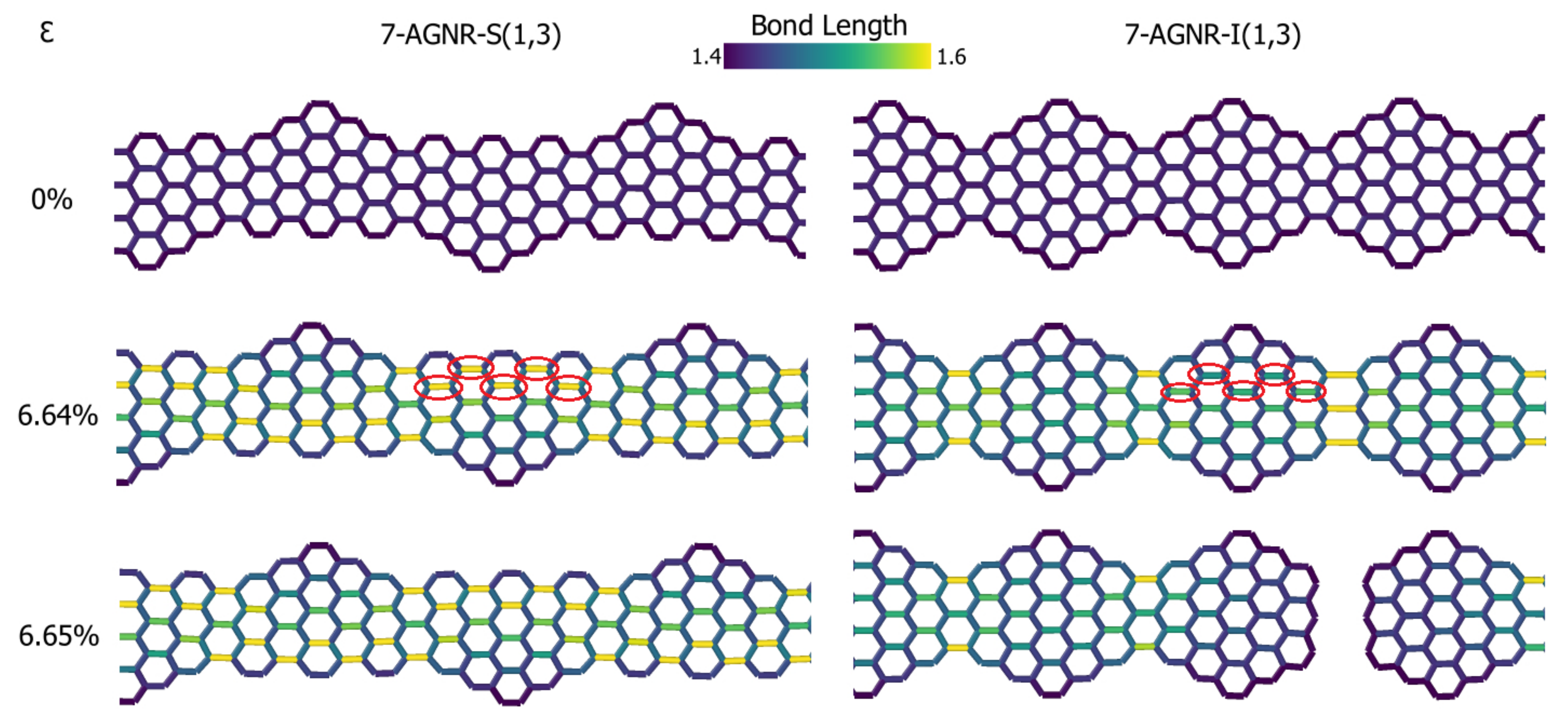}
    \caption{Snapshots showing deformation stages around the fracture strain for the 7-AGNR-$I$(1,3) superlattice, compared to the same strains for the 7-AGNR-$S$(1,3) superlattice. Some relevant spatially equivalent bonds are enclosed in red.}
    \label{fig:rupture}
\end{figure*}

The fracture mechanism always takes place in the thinner regions (\textit{i.e.} with the same width as the backbone) of the superlattice. In the Supplementary Material, the fracture mechanisms are displayed in detail for both modulated width AGNRs in Figures~\ref{fig:AGNRS} and ~\ref{fig:AGNRI}. Thinner regions allow stress concentration leading to bond fracture, while wider regions can distribute strain energy amongst more bonds. Figure~\ref{fig:rupture} allows us to visualize why the 7-AGNR-$I$(1,3) superlattice presents a fracture strain much lower than the 7-AGNR-$S$(1,3) case: the region between different segments generating modulation in the "inline" case allows for larger bond elongation leading to fracture. The "staggered case" allows for a different bond-strain distribution which defuses fracture, as shown by the bonds circled in red.
These simulation results indicate that modulated AGNR can withstand strains larger that those required in most technological applications, which are typically only a few percent.

\section{Electronic properties}
\label{sec:elec}

Before discussing the results of the superlattice under axial strain, let us summarize what is expected in $N$-AGNRs (\textit{i.e.} without the superlattice geometry). In AGNRs, the reciprocal space is just a one-dimensional projection that may or may not be commensurate with the position of the Dirac cone,  dividing the $N$-AGNRs into three groups: one metallic ($N=3p+2$), and two families of insulators ($N=3p, 3p+1$). Under axial strain, the Dirac cone of graphene moves apart from the $K(K')$ points, hence the distance of the different $k$-lines (of different $N$-AGNR families) to the Dirac cones is shifted too. Regarding the band gap under uniaxial strain, it increases for $N=3p+2$ and $N=3p$ AGNRs, while it decreases for $N=3p+1$ AGNR\cite{li2010strain}. For instance, without any strain the band gaps of 7- and 9-AGNR are $\Delta E_7\approx 1.38$ eV and $\Delta E_9\approx 0.92$ eV, respectively. As the 7- and 9-AGNR belong to different families ($3p+1$ and $3p$), it is expected that an axial strain will decrease $\Delta E_7$ but increase $\Delta E_9$. We calculated that under a strain of $\sim 3$\% both band gaps are the same $\Delta E_7=\Delta E_9=1.15$ eV. For a larger strain the larger band gap correspond to 9-AGNR ($\Delta E_9 > \Delta_7$). See Fig. \ref{fig:agnr_strain} for more details. This critical strain is much lower than the fracture strains found in \ref{sec:mech}. In the following, we will discuss the case of the interface between two $N$-AGNR under strain, and then the case of a superlattice.

To understand what to expect in a more complex AGNR,  we will make a couple of assumptions: (\textit{i}) the strain is uniform, and \textit{(ii)} the regions of the AGNR sections are long enough to have well-defined \textit{local} properties (\textit{e.g.} a local band gap). Let us start by considering the 7/9-AGNR interface, where a localized state appears. 
The edge state's penetration length ($\varepsilon$) through the 7- and 9-AGNR regions ($\varepsilon_7$ and $\varepsilon_9$, respectively) depends directly on the \textit{local} band gap of each region ($\Delta E_{7/9}$), and its analytical form\cite{note1} is $(e^{1/\varepsilon_{7/9}}-1)\propto \Delta E_{7/9}$. As the band gap of the 7/9-AGNR decreases/increases with uniform strain, the edge state's penetration length changes conversely. The interface of both AGNR systems has a critical point at a strain of $\sim 3$\%, where the band gap in both systems is the same and, consequently, the penetration lengths are also equal ($\varepsilon_7=\varepsilon_9$). Below this critical strain, the penetration length meets $\varepsilon_7<\varepsilon_9$, changing to a $\varepsilon_7>\varepsilon_9$ for values greater than 3\%. See Fig.~\ref{fig:kink} for more details of this process.

For a superlattice geometry, for example if we consider evenly spaced 7- and 9-AGNR regions, every interface state hybridizes with its corresponding adjacent interface state. In a low energy description, this hybridization can be considered a hopping between two interface states proportional to their penetration length, $t_{N}\propto e^{-L/\varepsilon_N}$, where $N=\{7,9\}$ and $L$ is the length of each region. The hopping between edge states is the foundation of the model at low energies, and an SSH-like effective Hamiltonian can be constructed with its different topological phases.\cite{munoz2018topological}  In the pristine case, for uniform strain up to 3\%, the band gap of the 9-AGRN is smaller, which means  $t_7>t_9$ in the low energy model. For the critical strain, $\sim 3$\%, both hopping strengths should be the same. Furthermore, for the last case, when the uniform strain exceeds 3\%, the situation is reversed, obtaining $t_7<t_9$. Interpreting this on the light of the SSH model, the transition in the hopping strength due to uniform strain implies a topological transition, resulting in new topological edge states corresponding to the superlattice geometry. It is worth remarking that we used some strong assumptions in this discussion, and the critical strain for the topological transition could be different. For instance,  Fig.~\ref{fig:rupture} shows non-uniform strain near fracture. However, the mechanism explaining the emergence of a topological transition should remain valid. 

\begin{figure}
    \centering
    \includegraphics[width=0.9\columnwidth]{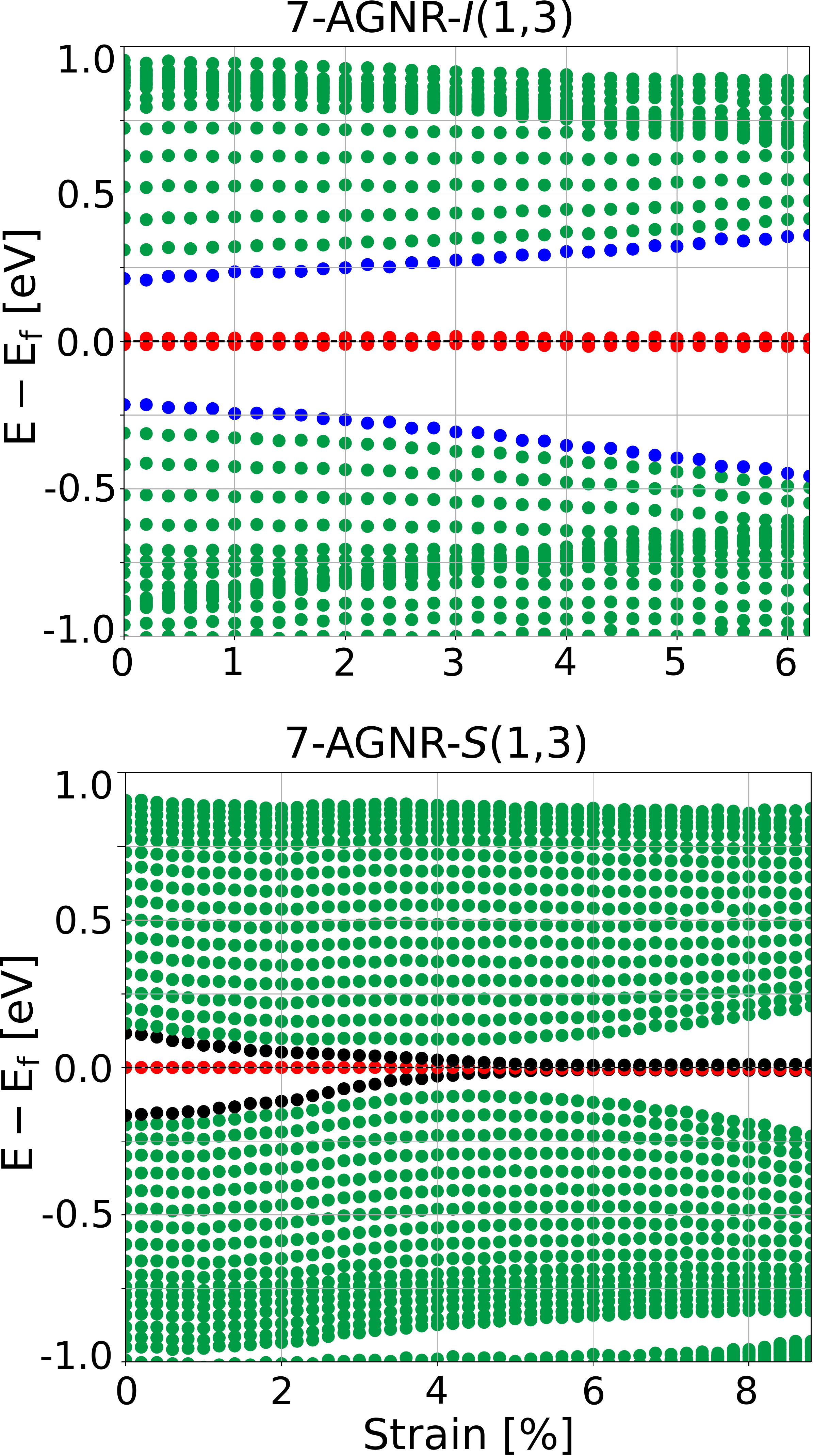}
    \caption{Evolution of the energy levels as a function of strain. The topological edge states are very close to $E=E_F$. In the 7-AGNR-$I$(1,3) there are four states colored red (two of them come from the zigzag borders). The 7-AGNR-$S$(1,3) superlattice has two topological states (colored red) from the zig-zag edges; meanwhile two bulk energy levels become topological at a strain close to 5\% (colored black). Two bands are colored blue only to make  Fig.~\ref{fig:sp} clearer. }
    \label{fig:bands-strain}
\end{figure}

For the electronic properties of the superlattice, we are interested in the region that precedes the fracture process. The evolution of the energy levels with the applied strain is shown in Fig.~\ref{fig:bands-strain}. All the levels shown are derived from the superlattice SSH-like states (the graphene bulk-bands are beyond this range). The first case under study is the 7-AGNR-$I$(1,3) superlattice. The system can be described as an SSH model in the topological phase for low energies, expecting two states in the Fermi energy. Further, two more edge states are contributed by the zig-zag termination in the edge.\cite{louis2019electron} The red dots in the upper panel of Fig.~\ref{fig:bands-strain} are the practically fourfold-degenerated edge states at the Fermi energy, and there is a small hybridization between zig-zag and topological edge states. 
As a non-zero uniform strain is applied in the edge states, the SSH bulk-like bands move away from the Fermi energy. This behavior in 7-AGNR-$I$(1,3) is expected due to lack of changes in the topological phase in the low energies SSH-like model, and it does not break any symmetry required for the appearance of these states, including the zig-zag edge states (at the end of the extra backbone).

The counterpart system, on the trivial phase without strain, is the 7-AGNR-$S$(1,3). The first substantial difference is the absence of two edge states in the Fermi energy due to this trivial phase. Without strain, the bottom panel in Fig.\ref{fig:bands-strain} shows the two-fold degenerated edge states coming from the zig-zag edge termination in the Fermi energy. Additionally, the green dots show trivial bands coming from SSH bulk-like states, and they approach or move away from the Fermi energy depending on whether the strain is less than or greater than $\sim4.5$\%. As with the previous case, the zig-zag edge states do not show significant changes when the strain increases. However, at a certain critical point (between 4\% and 5\%), two additional states begin to coexist at the Fermi level (black dots in the figure), and there is also an edge localization. The origin of these states can be explained as a change in the phase of the effective SSH model and originates due to the changes from one region to another (7- or 9-AGNR) of the hopping strength magnitude between adjacent boundaries states. This change from a trivial to the topological phase implies the emergence of topological states at the superlattice edge and remains very stable against the increase of strain.

\begin{figure*}
    \centering
    \includegraphics[width=\textwidth]{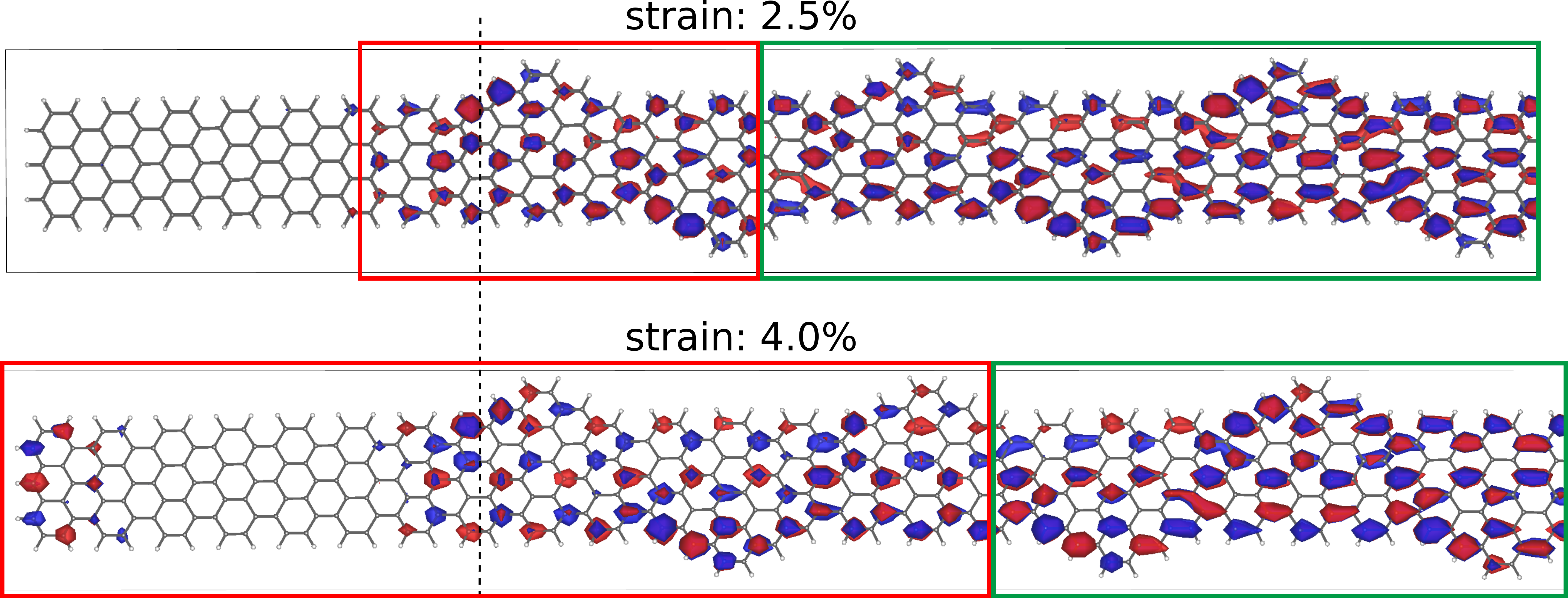}
    \caption{Topological transition of the 7-AGNR-$S$(1,3) under strain. The wave function of one of the levels involved (black symbols in Fig.
    ~\ref{fig:bands-strain}) is plotted. The regions of the band structure with/without an evident sublattice polarization (\textit{i.e.} as in a topological state) are enclosed by a red/green frame.}
    \label{fig:wf_transition}
\end{figure*}

\section{Sublattice Polarization and robustness of the edge states}
\label{sec:slp}

\subsection{Sublattice Polarization}
Generally, the system’s topology is described in the reciprocal space in the bulk system, whose consequences are observed in the edge. Nevertheless, recent studies have shown that real-space descriptors can characterize the topology of a system\cite{Huaqi18,troncoso20,martin2021under}. In particular, the sublattice polarization can characterize the topology of SSH-edge states, predicting the topological phase of the system regardless of bulk\cite{martin2021under}.  As we have seen previously, for low energies, AGNR superlattices under strain behaves like an SSH system, giving us the possibility to describe the topological states by means of the sublattice polarization. Indeed, the electrical transport in AGNR superlattices can be described by the degree of sublattice polarization of its topological-like modes.\cite{Rizzo2020sci} Before continuing, a few remarks about the sublattice polarization are needed. The definition of sublattice polarization, $|\psi_A|^2-|\psi_B|^2$, has been employed before\cite{pereira2008valley}, but it can not be readily applied for a superlattice geometry. The hybridization of the sublattice polarized states  (\textit{e.g.} $|\phi_A^{left}\rangle\pm|\phi_B^{right}\rangle$) gives a zero net polarization, regardless of the nature of the states.
The sublattice polarization is well preserved at a \textit{local} scale but not along the whole superlattice. We define a number $P_{A/B}$, that gives an account of the sublattice polarization of the whole system at the \textit{local} scale:

\begin{equation}
  P^n_{A/B} = \frac{1}{N}\left(\sum_{<i,j>} |\psi^{(n)}_i||\psi^{(n)}_j|\right)^{-1},
  \label{eq:sp}
\end{equation}

\noindent where $\psi^{(n)}$ is the wave-function of the $n$-th energy level, the summation runs over all nearest-neighbors pairs, and $N$ is the number of electrons. In a fully delocalized state without any sublattice polarization, $P_{A/B}\propto N^{-1}$. In a localized state without sublattice polarization (\textit{e.g} an H atom bonded to the AGNR), $P_{A/B}$ should be a small number since the summation is still close to one. However, in a perfectly sublattice polarized state, $P_{A/B}$ diverges, regardless of the localization. In practice, and with a superlattice geometry, a value of $P_{A/B}$ close or larger than one indicates sublattice polarization.

\subsection{Topological transition and robustness of the states}

\begin{figure}
    \centering
    \includegraphics[width=0.8\columnwidth]{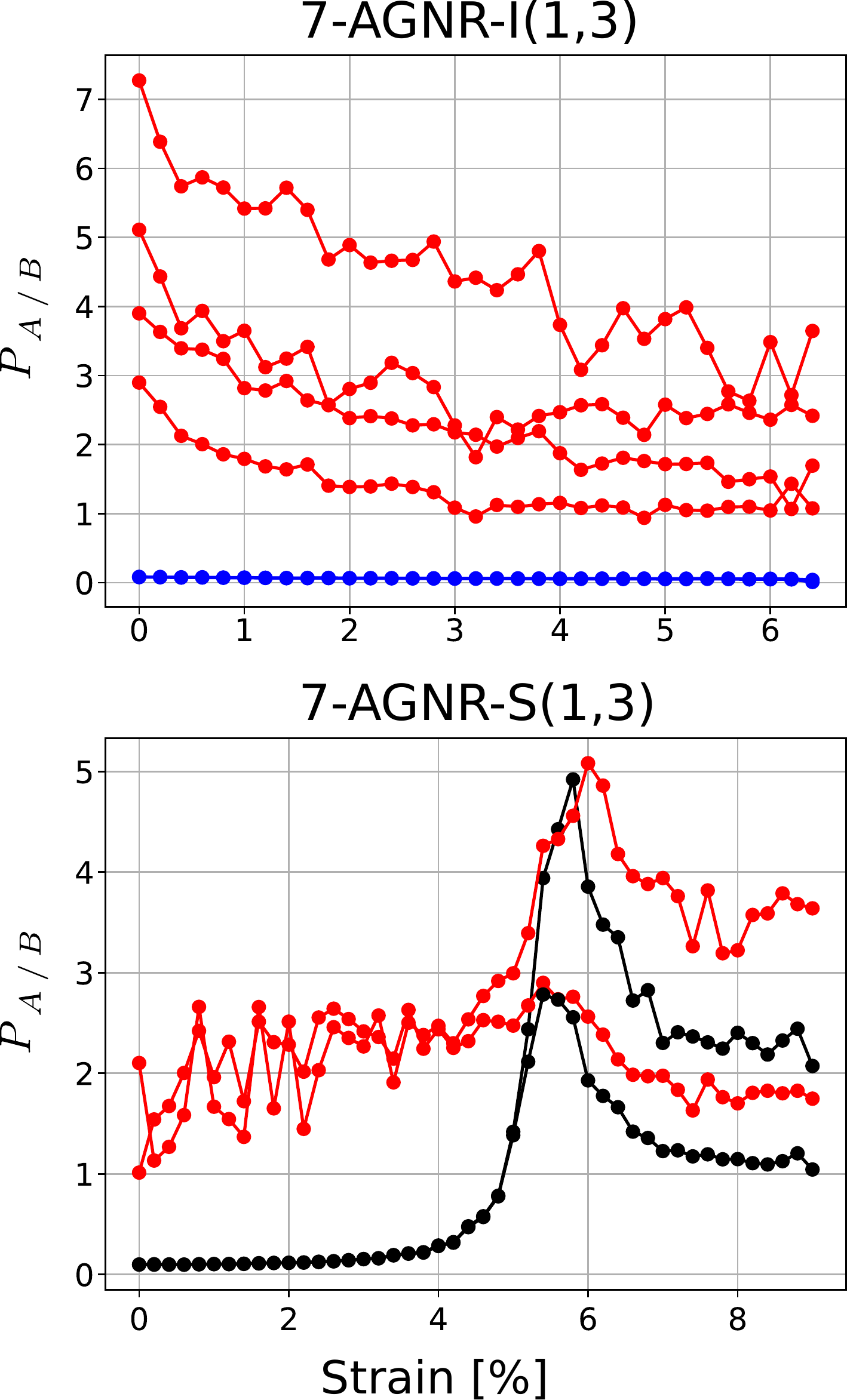}
    \caption{Sublattice polarization, as defined in Eq~\ref{eq:sp} for selected states of each superlattice. The states use the same color scheme of Fig.~\ref{fig:bands-strain}.}
    \label{fig:sp}
\end{figure}

Fig.~\ref{fig:sp} shows $P_{A/B}$ for some selected states of each superlattice. Conceptually, the case of 7-AGNR-$I$(1,3) is clear regardless of the strain, since the states colored blue (see Fig.~\ref{fig:bands-strain}) correspond to non-topological SSH-like energy levels, and they have $P_{A/B}$ close to zero. Instead, the four topological states -colored red- have  $P_{A/B}$ close to or larger than one. Each topological state has a different $P_{A/B}$, because the index is very sensitive to any local breaking of the sublattice polarization. For instance, when the topological states from the left and right borders hybridize, they form odd and even states. The even state slightly breaks the sublattice polarization at the center of inversion. However, the odd state does not (it has a node at the center), implying a consistently lower or larger $P_{A/B}$, according to their parity. Something similar happens with the hybridization between the zig-zag edge states and the ones at the superlattice's edges.

A remarkable topological transition near $5.0$\% can be seen in the bottom panel of fig.\ref{fig:bands-strain} for the 7-AGNR-$S$(1,3) case (black dots). Below the topological transition, $P_{A/B}$ increases slowly with the strain, from its bulk value, but near the transition its increase is very rapid, suggesting the utility of this index to characterize the process. 

There is another qualitative change induced by strain, and in the discussion below we will argue its relationship with an enhanced robustness of the topological states due to the superlattice. The topological zig-zag edge states (red dots in the bottom panel of Fig.~\ref{fig:sp}) present a noisy sublattice polarization for strain values below the topological transition of the superlattice. However, the noise is practically suppressed after the transition. Even though the topologically-protected edge states are robust to the disorder caused by the dynamics (on symmetry grounds and also because \textit{i.e.} $P_{A/B}\geq 1$), they are not impervious to that disorder, and the noise on $P_{A/B}$ reflects changes in the wave function (\textit{e.g.} its distribution in real space can change). After the transition, two SSH-like edge states appear, hybridized with the zig-zag states. These states are quite peculiar, because they inherit properties from the superlattice geometry (SSH) and from the underlying graphene-like lattice.

The localization length of the superlattice edge states has a different scale compared to regular zig-zag edge states. In the former the typical length scale are superlattice cells (see Fig.~\ref{fig:scheme}), while in the latter it is the graphene unit cell. This difference in the localization implies that the superlattice states are \textit{specially} robust to disorder in the underlying hexagonal lattice: they experience the average disorder over several sites, which tends to cancel, in stark contrast with the disorder at the atomic scale. This mechanism has been explained in detail in other related one-dimensional superlattices.\cite{munoz2018topological}.

This \textit{extra} resilience of the topologically protected superlattice states does not change anything fundamental: the states from both edges are to be degenerate within the band gap, as it happens with regular zig-zag edges. However, this property can be very useful for applications involving the interaction of these edge states; an edge state mostly insensitive to disorder would imply a steady interaction, even with a large temperature. For instance, among the first applications of these AGNR superlattices are simulators of one-dimensional Heisenberg antiferromagnets \cite{Cao2017Topo}. This effective magnetic interaction is created by the overlap of superlattice states with atomic defects, and a steady overlap would be beneficial to overcome the effects of temperature.

\section{Methods}
\label{sec:methods}

For calculating the mechanical properties of AGNRs, we used classical molecular dynamics (MD) simulations , applying uniaxial deformation to the AGNR. MD simulations were carried out with the LAMMPS software\cite{plimpton1995}, using the REBO-scr potential.\cite{pastewka2008} It has been shown that this potential gives a reliable description of the strain-stress processes (including bond breaking) in AGNRs and other C-based systems\cite{aparicio2020simulated,tangarife2019molecular}. Studied superlattices have finite size and were thermalized at 10 K. We used long periodic superlattices, and homogeneous strain was applied, with a strain rate of $1\times 10^{9}$ s$^{-1}$.  Visualization was done with OVITO \cite{stukowski2009}. See Supplementary Material for further details.

The electronic structure was calculated using some of the atomic configurations obtained by the MD simulations explained above. Given the large number of electrons in the superlattices, most of the calculations were made with density functional based tight-binding, as implemented in the DFTB+ package.\cite{hourahine2020dftb+,mio} We compared the electronic structure from DFTB+ with the results from  the density functional theory package VASP,\cite{vasp1,vasp2,vasp3,paw} with the PBE exchange-correlation,\cite{pbe} obtaining good agreement. This is not surprising, since even a simpler tight-binding with a fixed hopping parameter has good agreement with experiments (without strain).\cite{groning2018} Also, our calculations of the phonon spectrum of graphene with DFTB and DFT gave similar results. For the analysis of the electronic structure we used a modified version of PyProcar\cite{pyprocar} that exploits the sublattice polarization of the topological states.

\section{Summary and Conclusions}

We studied the electronic and mechanical properties under strain of superlattices formed by 7-AGNR and 9-AGNR. These AGNR have a different geometrical phase, and the entire superlattice -at low energies- resembles the SSH model: a bipartite chain with nearest neighbors interactions $t, t'$. The model has two topological phases: $t>t'$ and $t<t'$, and depending on the actual edges one of phases has zero-energy topologically-protected edges. 

Our main contribution is to identify a mechanism to induce a topological transition by means of strain. Different topological classes of AGNR, such as 7-AGNR and 9-AGNR, have an opposite evolution of the band gap with strain. This implies an opposite evolution of the localization length of the boundary states, \textit{i.e.} the states forming the SSH-like system. Therefore, the interactions $t,t'$ have an opposite behavior with respect to strain. For the particular case of the superlattices formed by 7-AGNR and 9-AGNR, the interaction through the 7-AGNR increases with strain, but the interaction passing through the 9-AGNR decreases with strain. Depending on what interaction is larger without any strain, a topological transition should take place.

The topological states of the superlattice -if present- are robust to strain even close to fracture. This is not surprising since strain does not breaks the chiral symmetry protecting the edge states. Nevertheless, by means of the sublattice polarization, we found that the superlattice edge states are very stable in real space. This is not the case for the common edge states of zig-zag GNR. This could be useful to overcome thermal fluctuations in the implementation of artificial systems using these edge states as building blocks.

Classical molecular dynamics simulations show that mechanical properties change due to modulation, but staggered graphene nanoribbons have nearly the same fracture strain as unmodulated ribbons, therefore they support a large strain and this mechanism of topological transition should be valid for several other types of superlattices.

%Despite the large amount of research about controlling the electronic structure in AGNRs, it has been difficult to control its topological properties. This is because the typical band gaps of AGRN is in the range of $\sim 1$ eV, and a topological transition needs to overcome (close) that band gap. By using a superlattice geometry, this obstacle was eliminated, and the band gap can be designed by the superlattice geometry\cite{rizzo2018,groning2018}.  

Recently, two specific GNR superlattices have been synthesized resembling massive\cite{Sun20adma} or even massless\cite{Rizzo2020sci} Dirac fermions at the low-energy limit. Here, by using a moderate strain in 7-AGNR-S(1,3), we found a continuum of band gaps, including both the massless and massive Dirac fermion behaviors, close and just above the transition, respectively. 
The superlattice geometry of AGNR explored here opens a way to induce several exotic (mostly topological) effects, which could have applications in electronics and quantum computing.

\begin{acknowledgments}
E.A. and EMB thank support from ANPCyT grant PICTO-UUMM-2019-00048, and SIIP-UNCuyo grant 06/M104. This work was partially supported by Fondecyt Grants No. 1191353, 11180557, 11190484, 1190662; J.D.M. was funded by the National Agency of Research and Development (ANID) through grants Fondecyt posdoctorado number 3200697; Center for the Development of Nanoscience and Nanotechnology CEDENNA AFB180001 and from Conicyt PIA/Anillo ACT192023. This research was partially supported by the supercomputing infrastructure of the NLHPC (ECM-02).
\end{acknowledgments}

%\section{SUPPLEMENTARY MATERIAL}

\bibliographystyle{apsrev4-1_title}
\bibliography{bib}% Produces the bibliography via BibTeX.

\begin{thebibliography}{57}
\bibitem [{not()}]{note1}%
  \BibitemOpen
  \href@noop {} {\enquote {\bibinfo {title} {The proportionality factor can be
  different for different {N-AGNR}, but we will ignore it during the
  discussion, the most important part is its dependency with the band gap: The
  localization length decreases (increases) as the band gap increases
  (decreases)},}\ }\BibitemShut {NoStop}%
\end{thebibliography}

%%%%%%%%%% Merge with supplemental materials %%%%%%%%%%
\newpage
\widetext
\begin{center}
\textbf{\large Supplemental Materials \\ Inducing a topological transition in graphene nanoribbons superlattices by external strain \\ Topological transition in graphene nanoribbon superlattices induced by external strain}
\end{center}
%%%%%%%%%% Merge with supplemental materials %%%%%%%%%%
%%%%%%%%%% Prefix a "S" to all equations, figures, tables and reset the counter %%%%%%%%%%
\setcounter{equation}{0}
\setcounter{figure}{0}
\setcounter{table}{0}
\setcounter{page}{1}
\makeatletter
\renewcommand{\theequation}{S\arabic{equation}}
\renewcommand{\thefigure}{S\arabic{figure}}
\renewcommand{\bibnumfmt}[1]{[S#1]}
\renewcommand{\citenumfont}[1]{S#1}
%%%%%%%%%% Prefix a "S" to all equations, figures, tables and reset the counter %%%%%%%%%%

\section{Fracture Process}

\begin{figure}[h]
    \centering
    \includegraphics[width=0.3\columnwidth]{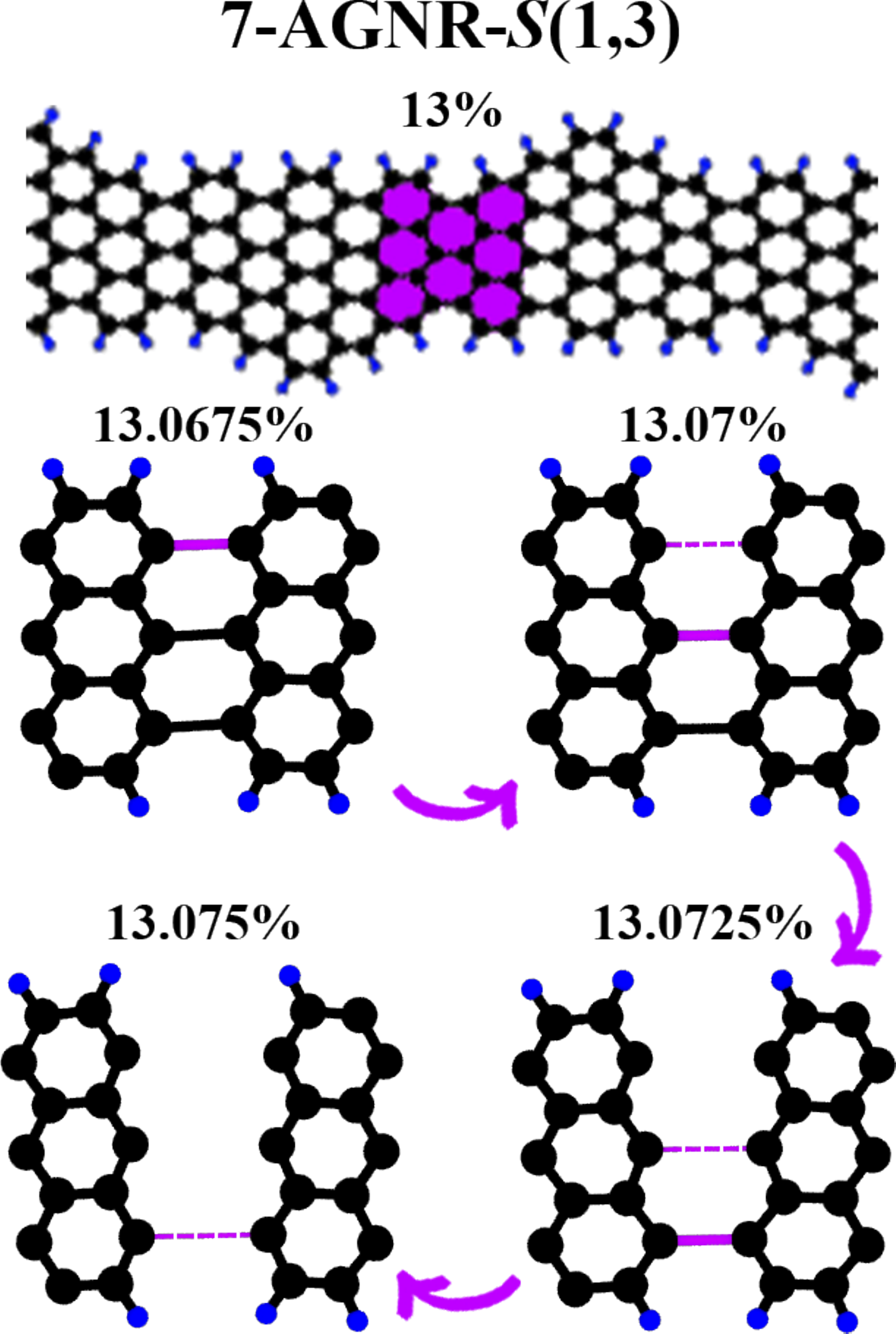}
    \caption{7-AGNR-$S$(1,3) superlattice at 13\% strain when no bond has broken yet. Carbon atoms in black and hydrogen atoms in blue. The purple region marks the zone that will be involved in the fracture process. Detailed evolution of bond breaking in that purple region is presented as a function of strain, where purple thick bonds will break in the next snapshot and will be represented by dashed thin purple bonds.}
    \label{fig:AGNRS}
\end{figure}

\begin{figure}[h]
    \centering
    \includegraphics[width=0.3\columnwidth]{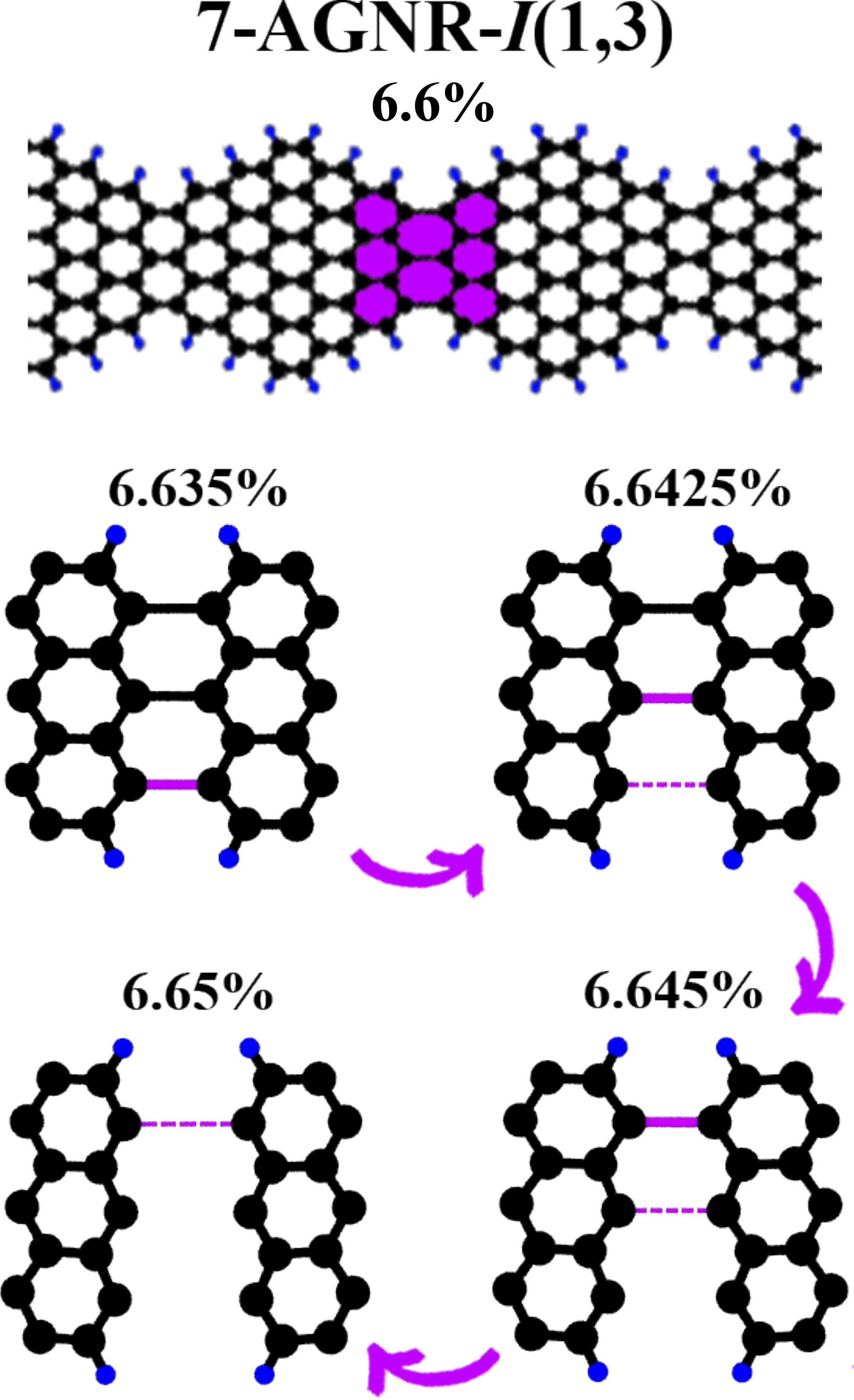}
    \caption{7-AGNR-$I$(1,3) superlattice at 6.6\% strain when no bond has broken yet. Carbon atoms in black and hydrogen atoms in blue. The purple region marks the zone that will be involved in the fracture process. Detailed evolution of bond breaking in that purple region is presented as a function of strain, where purple thick bonds will break in the next snapshot and will be represented by dashed thin purple bonds.}
    \label{fig:AGNRI}
\end{figure}

In order to better understand the modulated AGNR fracture process, we include some MD snapshots in Figures~\ref{fig:AGNRS} and ~\ref{fig:AGNRI}.

\section{Electronic Properties}

\begin{figure}[h]
    \centering
    \includegraphics[width=0.7\textwidth]{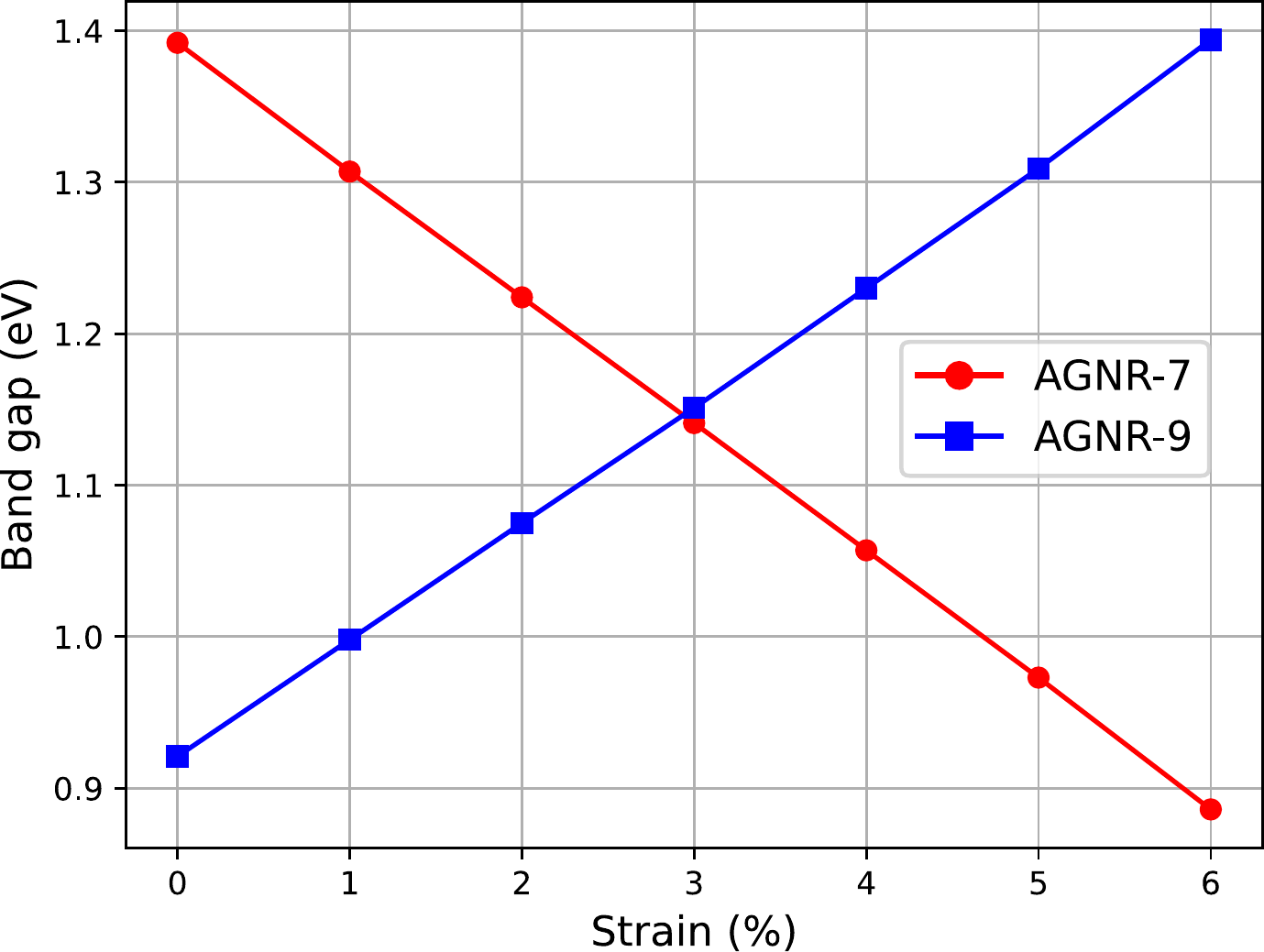}
    \caption{Evolution of the band gap as a function of strain. The AGNRs are relaxed (the only constraint is the strain). The 7-AGNR and 9-AGNR belong to different topological phases, hence their band gaps have the opposite behavior with strain.}
    \label{fig:agnr_strain}
\end{figure}

\begin{figure}[h]
    \centering
    \includegraphics[width=\textwidth]{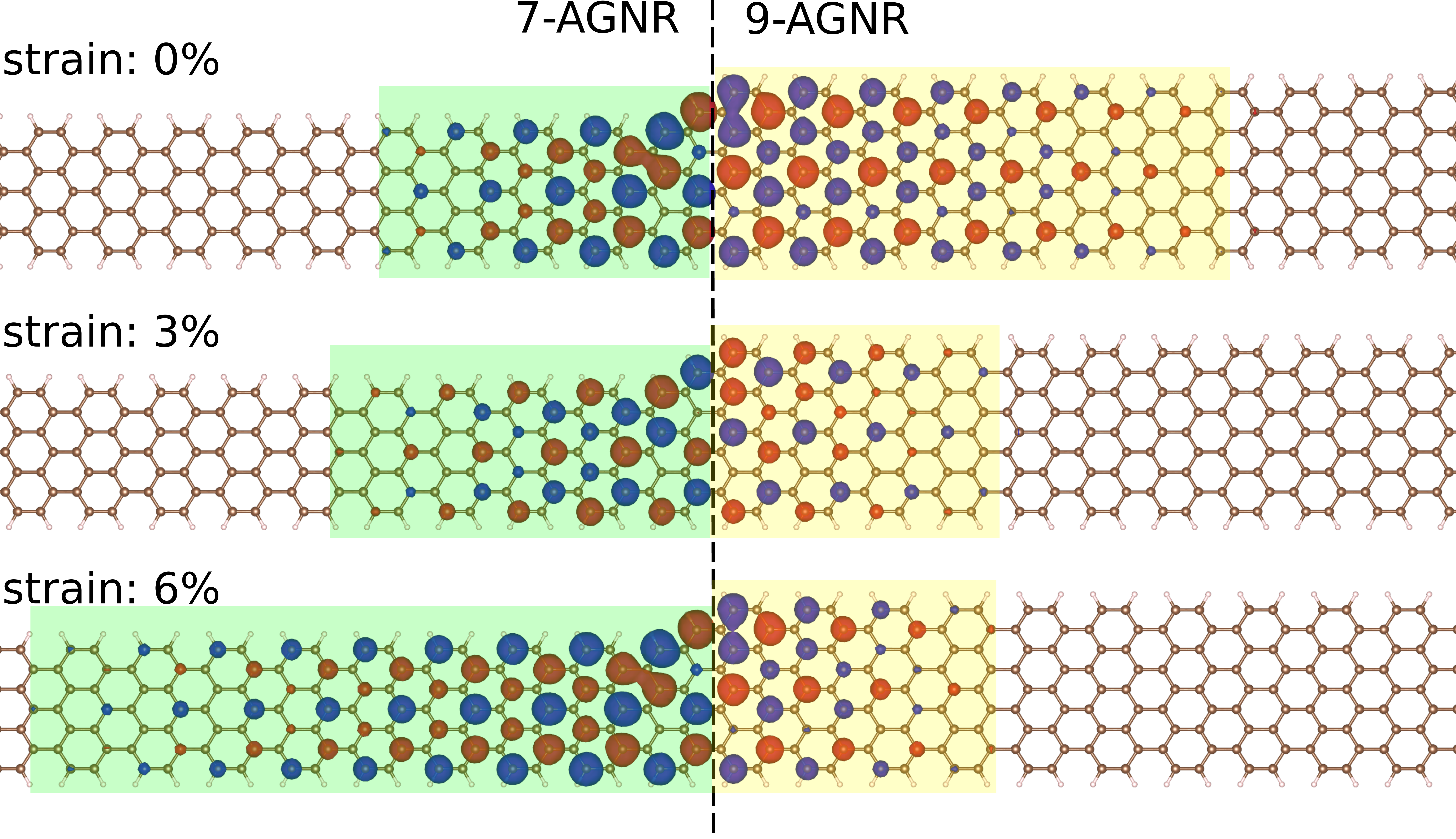}
    \caption{Localization of the interface wave function as a function of the applied strain. There is a clear correspondence between the \textit{local} band gap of the N-AGNR and the localization of the electronic state along each segment: the smaller the band gap the larger the localization. To avoid a non-uniform strain along the sample only the ground-state structure was relaxed. The isovalue is 0.005, and the shaded regions are just a guide to the eye.}
    \label{fig:kink}
\end{figure}

\end{document}